\documentstyle[aps,epsf]{revtex}
\def\nn {\nonumber}

\def\e{{\rm e}}

\def\d{{\rm d}}
 \def\(({\left(}
 \def\)){\right)}

\def \d{{\rm d}}
\def \beq{\begin{equation}}
\def \eeq{\end{equation}}
\def \ln{{\rm ln}}

\def \ln{{\rm ln}}

\def \ab2{\alpha\beta^2}

 \newcommand {\be} {\begin{equation}}
\newcommand {\bea} {\begin{eqnarray} \nonumber }
\newcommand {\ee} {\end{equation}}
\newcommand {\eea} {\end{eqnarray}}
 \newcommand {\eps} {\epsilon}

\newcommand {\si} {\sigma}

\input{epsf}
\begin{document}
\title{Non trivial overlap distributions at zero temperature}

\author{ Silvio Franz$^{1}$, Giorgio Parisi$^{2}$\\
}
\address{
$^{1}$The
Abdus Salam International Center for Theoretical Physics\\
Strada Costiera 11,
P.O. Box 563,
34100 Trieste (Italy)\\
$^{2}$ Dipartimento di Fisica, Sezione INFN and Unit\`a INFM \\
Piazzale Aldo Moro 2, 00185, Roma (Italy)
}
\date{\today}
\maketitle

\begin{abstract}
We explore the consequences of Replica Symmetry Breaking at zero
temperature. We introduce a 
repulsive coupling between a system and its unperturbed ground
state. In the Replica Symmetry Breaking scenario a finite coupling
induces a non trivial overlap probability distribution among the
unperturbed ground state and the one in presence of the coupling. 
We
find a closed formula for this probability for arbitrary
ultrametric trees, in terms of the parameters defining the tree.
The
same probability is computed in numerical simulations of a simple
model with many ground states, but no ultrametricity: polymers in
random media in 1+1 dimension. This gives us an idea of what violation
of our formula can be expected in cases when ultrametricity does not hold. 

\end{abstract}

\section*{}

\section{Introduction}

In recent times there has been a wide interest in the behavior spin glasses with Gaussian
couplings at zero
temperature.

Some of the reasons for this interest are the following:
\begin{itemize}
    \item  The energies are continuous variables and the ground state is unique. It is also natural
    to  suppose (although it is far from being proved) that the limits $T \to 0$ and $N \to \infty$
    do commute and therefore the shape of the energy landscape is similar to that of the free
    energy landscape at non-zero temperature (for a discussion of this point see \cite{MMPRRZ}).

    \item  Working at zero temperature avoids completely the possibility that the temperature
    used is too near to the critical point.

    \item  Technical progresses has been done in the algorithm for finding the ground state
    \cite{PY,HM,KM,MP,PAL}
    and it is
    now possible to studies three dimensional systems up to $14^{3}$ spins \cite{MP}.
\end{itemize}

In this framework it has been suggested 
 that 
a possible test of the
applicability of the Replica Symmetry Breaking (RSB) scenario
is the study
of the overlap of the ground state of two systems whose
total Hamiltonian differs by a quantity of
order 1
 \cite{PY2,KM}. 

Let us consider a simple case. We have a first
system with Hamiltonian $H_{0}({\bf\si})$ and its
ground state is given by $\tau_i$.
We now consider a second system whose Hamiltonian is
\be
H_{1}({\bf\si})=H_{0}({\bf\si})+\eps H_{{\bf\tau}}(\si)  \  .
\ee

Three quite simple choices of $H_{{\bf\tau}}({\bf\si})$ are:
\be
H_{{\bf\tau}}({\bf\si})=q(\si,\tau),
\ \ \ \ H_{{\bf\tau}}({\bf\si})=q^{2}(\si,\tau), \ \ \
\ H_{{\bf\tau}}({\bf\si})=q_{l}(\si,\tau)   \  ,
\ee 
where the overlap $q$ and the link
overlap $q_{l}$ are given by 
\bea q(\si,\tau)= N^{-1}\sum_{i}\si_i\tau_i  \  ,\\
q_{l}(\si,\tau)=N_{l}^{-1}\sum_{\langle i,k\rangle}\si_i\tau_i\si_k\tau_k  \  , 
\eea
where the sum is done over all the nearest neighbor pairs $(i,k)$ in a
short range model or over all the pairs in the SK model ($N_{l}$ being
the total number of pairs in this sum). The third possibility has been 
actually used by \cite{PY2}.

In presence of quenched disorder, the the value of the
overlap among the ground states of $H_0$ and $H_1$ can be sample
dependent. 
This observation can be used tp  the starting point for
 investigating possible RSB in the three dimensional Edwards-Anderson
model. 
The question we address in this paper, is the computation of the probability
distribution induced by the random couplings of $q$ or of $q_{l}$
among the two ground states, in the hierarchical RSB scenario.

Obviously the choice $H_{q}(\si)=q$ is interesting only in presence of
a magnetic field which breaks the symmetry $\si(i)\to - \si(i)$,
otherwise we would get that $\si(i)=-\tau(i)$ for positive non zero
$\eps$ and $q=-1$. The second choice is more interesting at zero
magnetic field, but it is slightly harder to implement numerically,
because its non local nature. The third choice is however equivalent
to the first one in the SK model, where is known that $q_{l}=q^{2}$
apart from corrections that vanishes when the number of spins goes to
infinity. In short range models, it is possible (as suggested by the
principle of replica equivalence \cite{repequiv}) that with probability one when the
volume goes to infinity $q_{l}=f(q^{2})$, where $f$ is a function that
can be determined numerically and which should be not too far from
\be
f(q^{2})=A+(1-A)q^{2} \ .
\ee

It is evident that for finite $\eps$ the perturbation is of order 1
and it quite interesting that if replica symmetry is broken the
function $P(q)$ is non trivial at $\eps \ne 0$. Let us define as
$E_{gs}(0)+\Delta(q)$ the energy of the first excited state of the
Hamiltonian $H_0$ with an overlap $q$ with the ground state. The
ground state of the Hamiltonian $H_1$ is
\begin{equation}
E_{gs}(\eps)=E_{gs}(0)+\min_{0\leq q\leq 1} \{ \Delta(q)+\eps g(q)\}
\end{equation}
where $g(q)=q,q^2,q_l$. 
The main achievement of this paper will be the computation in section
II and III of the joint probability distribution of $\Delta$ and $q$
for which the minimum is attained, for arbitrary RSB trees. In section 
IV we will give some example in mean field models, 
while in section V we show the result of a numerical 
computation for directed polymers in random media in 1+1 dimension.

\section{Replica symmetry breaking}

The computation presented in this note could
be done in two different ways:
\begin{itemize}
    \item Using the replica formalism \cite{MPV} to compute the partition
    function of the perturbed Hamiltonian $H_{1}$.

\item 
Exploiting
    directly the information coming from replica symmetry breaking on
    the probability distribution of the lowest lying states and doing
    a pure probabilistic computation.
\end{itemize}
The first alternative leads to some apparently messy combinatorial
analysis so that we have decided to follow the second alternative. In
this case the computation is physically instructive.

In this section we will recall, without proof, some known results
about replica symmetry breaking and also find some new consequences of
those results. Let us assume that replica symmetry is broken in the
system we consider and its breaking is  characterized by a function
$x(q,T)$ such that in the low temperature limit
\be x(q,T) =T
y(q)+O(T^{2})  \  ,
\ee
where the function $y(q)$ may be singular at $q=1$
(in the SK model it diverges as $(1-q)^{-1/2}$ near $q=1$) \cite{PaT}.

The space of lower lying configurations  is organized in a
rather complex way.

\subsection{One step replica symmetry breaking}

In this case only two values of the overlap are allowed ($q_{0}$ and
1), i.e. all different minima have a mutual overlap equal to
$q_{0}$. If we call $R$ a reference total energy, which depends on the
choice of the systems, i.e. on the variables $J$, the probability to
find a configuration in the interval $(E,E+dE)$ is given by
\be
\nu_0(E|R)\equiv \exp(y_{0}(E-R))  \  .
\ee  
We notice that configurations which differs by a
number of spin flips which remains finite when the volume goes to
infinity are identified.

This well known results has the consequence that the probability
distribution of the ground state energy $E_{0}$ is given by the
Gumbel law
\be \mu_0(E_{0}|R)=\exp(y_{0}(E_{0}-R))\exp
(-A_{0}\exp(y_{0}(E_{0}-R))  \  ,
\ee
with $A_0=1/y_0$. This formula is easily understood noticing that
the probability that there
are no configurations for $E'<E$ is given by \be
\exp\left(-\int_{-\infty}^{E}dE' \exp(y_{0}(E'-R))\right) = \exp
(-y_{0}^{-1}\exp(y_{0}(E-R))   \  .\ee

In the same way we obtaining that the probability of having a ground state at $E_{0}$ and
the first excited configuration at $E_{1}$ is given by:
\begin{eqnarray}
P(E_{0},E_{1}|R)& =& \nu_0(E_0|R)\mu_0(E_1|R)\nn  \  ,\\
& =& \exp(y_{0}(E_{0}-R))\exp(y_{0}(E_{1}-R))\exp (-y_{0}^{-1}\exp(y_{0}(E_{1}-R))  \  .
\end{eqnarray}

Finally if we define $\Delta_{0}=E_{1}-E_{0}$, the probability distribution of
$\Delta_{0}$, integrated over $E_{0}$ and $E_{1}$, one finds the simple result
\be
P(\Delta_{0})=y_{0}^{-1}\exp( -y_{0} \Delta_{0})  \  ,
\ee
for positive $\Delta_{0}$, the probability being obviously zero for
negative $\Delta_{0}$.

\subsection{Many level replica symmetry breaking}

Let us in this section generalize the computation for an arbitrary RSB
tree (see figure). As customary, we will first consider a tree with
$k$ levels and at the end we will generalize the result to the
continuous branching limit. The construction of the tree has been
described many times, and we only repeat it briefly to fix the
notation. At each node of the tree at the level $l$ it is assigned an
energy $E_l$, which is chosen in such a way that the  number of
nodes with energy in the interval $(E_l,E_l+dE_l)$ branching from a
node with energy $E_{l-1}$ is a Poisson variable with average
equal to $\exp(y_l(E_l-E_{l-1}))\d E_l$. We will consider the
case in which for all $l$, $y_{l+1}>y_l$, which will be a necessary
condition of convergence of the integrals that appear in the
computation.
The root energy
 $E_{0}=R$ is the
reference energy of the previous section. We call $l$-clusters,
the set of
branches which coincide at the $l$-th level
of the tree.

\begin{figure}
\epsfxsize=200pt
 \epsffile{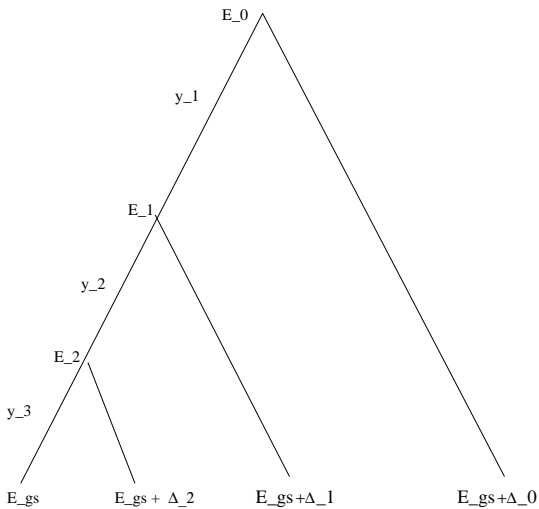}
\caption[0]{\protect\label{fig1} A three level tree, 
expliciting the notation we use in the text. }
\end{figure}

Let us define for any $l=k-1,...,1$ the first $l$-excited state, as
the first excited state which is in the same $l$-cluster as the ground
state, but in a different $l+1$-cluster, and denote its energy
$E_{gs}+\Delta_l$. 
In this section  we 
compute
the joint probability distribution of all the $l$-gaps $\Delta_l$.
We will get this quantity by first computing
$P_k(E_{gs},E_{gs}+\Delta_k,...,E_{gs}+\Delta_1|E_0)$ and then integrating
over $E_{gs}$.  
In that computation we make use
of the
following properties:

\begin{enumerate}

\item
The probability of a ground state $\mu_l(E|E_l)$ of an $l$-cluster (we
call it an $l$ ground state) is given by
\begin{equation}
\mu_l(E|E_l)=\exp\left( y_{l+1}(E-E_l)+A_l \e^{y_{l+1}(E-E_l)}\right) \ .
\label{mu}
\end{equation}
where the $A_l$ are a positive constants whose value could be easily
compute, but we will not need.  For $l=k$ the formula was derived in
the previous section.  Let us now proceed by induction supposing that
the formula holds for $l+1$ and show that it holds for $l$.  
Under the
induction hypothesis, we find that the number of $l+1$-ground states
with energy $E$ in an $l$ cluster is given by
\begin{eqnarray}
\nu_{l+1}(E|E_l)\d E_l & =& \int d E_{l+1} \;
\e^{y_{l+1}(E_{l+1}-E_l)} \mu_{l+1}(E|E_{l+1})
\nonumber\\
& = & const\times \e^{y_{l+1}(E-E_l)} \d E_l
\end{eqnarray}
from which we immediately find that the distribution of the $l$ ground
state is given by
(\ref{mu}) exploiting the reasonings of the previous sub-section.

\item
The joint probability
$P_k(E_{gs},E_{gs}+\Delta_k,...,E_{gs}+\Delta_1|
E_k,E_{k-1},...,E_0)$
can be written as:
\begin{eqnarray}
& &P_k(E_{gs},E_{gs}+\Delta_{k-1},...,E_{gs}+\Delta_0|
E_{k-1},...,E_0)=\nonumber\\
& &\nu_k(E_{gs}|E_{k-1})
\mu_k(E_{gs}
+\Delta_{k-1}|E_{k-1})
\mu_{k-1}(E_{gs}+\Delta_{k-2}|E_{k-2})
\times...\times\mu_1(E_{gs}+\Delta_0|E_0)
\end{eqnarray}
from which we get:
\begin{eqnarray}
& &P_k(E_{gs},E_{gs}+\Delta_k,...,E_{gs}+\Delta_0|E_0)= \nn\\
& & \int \d
E_{k-1}...\d E_1\;
\nu_k(E_{gs}|E_{k-1})\mu_k(E_{gs}+\Delta_k|E_{k-1})
\e^{y_{k-1}(E_{k-1}-E_{k-2})}
\mu_{k-1}(E_{gs}+\Delta_{k-2}|E_{k-2})\e^{y_{k-2}(E_{k-2}-E_{k-3})}
\times\nn\\
& &...\times\mu_{2}(E_{gs}+\Delta_{2}|E_{1})\e^{y_{1}(E_{1}-E_{0})}
\mu_{1}(E_{gs}+\Delta_{1}|E_{0})
\end{eqnarray}

\item
A detailed computation shows that
\begin{equation}
\int \d
E_{k-1}\;
\nu_k(E_{gs}|E_{k-1})\mu_k(E_{gs}+\Delta_k|E_{k-1})
\e^{y_{k-1}(E_{k-1}-E_{k-2})}=(y_k-y_{k-1})\e^{-(y_k-y_{k-1})\Delta_k}
\nu_{k-1}(E_{gs}|E_{k-2}).
\end{equation}
This allows to integrate
all the $E_l$ ($l=k-1,...,1$) telescopically, and
obtain
\begin{equation}
P_k(E_{gs},E_{gs}+\Delta_k,...,E_{gs}+\Delta_1|E_0)=
\prod_{i=2}^k \left( (y_i-y_{i-1})  \e^{-(y_i-y_{i-1})\Delta_i}\right)
\nu_1(E_{gs}|E_0)\mu_1(E_{gs}+\Delta_1|E_0)
\end{equation}

\item
We can finally integrate $E_{gs}$ as in the previous section
and get that the gaps' probability distribution is:
\begin{equation}
P(\Delta_k,...,\Delta_1)=
\prod_{i=1}^k \left( (y_i-y_{i-1})  \e^{-(y_i-y_{i-1})\Delta_i}\right)
\label{discreta}
\end{equation}
having defined $y_0=0$.
\end{enumerate}

We can now consider the continuum branching limit, in which the
branches can be indexed by the value of q, or by any monotonically
increasing function of $q$.
$y_i\to y(q)$, $y_i-y_{i-1} \to y'(q)\; dq$
where the previous formula
reduces to
\begin{equation}
P(\{\Delta (q)\})=  \exp\left(-\int_0^1 dq
y'(q)\Delta(q)\right)\prod_{q=0}^1 y'(q)\; dq \ ,
\label{continua}
\end{equation}
where, if $y_1\to y(0)\ne 0$, we make the convention that
$y'(0)=y(0)\delta(q)$.
\section{The overlap probability distribution}

We are finally in the position to compute the joint distribution
of the gap and of the overlap.
For a given sample, as we said,  the difference among the ground state energy
of the Hamiltonians $H_1$ and $H_0$
is given by the
\begin{equation}
E_{gs}(\epsilon)-E_{gs}(0)=\min_{0\leq q\leq 1} \Delta(q)+\epsilon g(q). \label{ga}
\end{equation}
Noticing that the indexing of the tree could be done by the function
$g(q)$ itself,
we can concentrate here to case $g(q)=|q|$, and consider only
positive overlaps. All the other cases can be obtained by this one
via a simple change of variable.
Let us call $\Delta$ and $q$ the arguments of
the minimum in (\ref{ga}).
Notice that if we want $q\ne 1$, $\Delta$  has to verify the
inequality
$\Delta\leq \epsilon (1-q)$, which expresses the fact that $\Delta(1)$
is always equal to zero.

We get the probability $P(\Delta,q)$  integrating
in formula  (\ref{continua}) all the $\Delta(q')$ ($q'\neq q$) with the
condition $\Delta(q')\geq \min\{0,\Delta+\epsilon (q-q')\}$,
so that, if $q'>\Delta/\epsilon+q$ the integration over $\Delta(q')$
will contribute with a one, while, in the opposite case
$q'<\Delta/\epsilon+q$ it contributes with the factor $\exp(-\epsilon dq'\;
y'(q')(\Delta/\epsilon+q-q'))$. We finally get:

\begin{equation}
P(\Delta, q)=
\theta(1-q-\Delta/\epsilon) y'(q)\exp\left(-\epsilon \int_0^{q+\Delta/\epsilon}
d q'\; (y(q')-y(0))\right)+\delta(\Delta)\delta(q-1)\exp (-\epsilon \chi) \ ,
\end{equation}
where we defined $\chi=y(0)+\int_0^1 dq\; y(q)$. The factor $y'(q)$ that
multiply the exponential comes from the only $\Delta$ which has
remained unintegrated.

Let us notice that the formula depends on $\Delta$ only in the
combination $\Delta/\epsilon+q$.

Integrating over $\Delta$ we get, for the overlap probability
\begin{equation}
P(q)=\delta(q-1)\exp(-\epsilon\chi)+\epsilon y'(q)\int_q^1 dq'
\exp\left(-\epsilon
\int_0^{q'}dq''(y(q'')-y(0))\right)
\label{P}
\end{equation}

It is also interesting to study the probability distribution of
$w=\Delta/\eps+q$. Integrating over $\Delta$ and $q$ for fixed $w$
with the condition $0\leq q\leq w$ we find the remarkable formula:
\begin{equation}
P(w)=\theta (1-w) \eps (y(w)-y(0)) \exp\left(-\eps\int_0^w \d q \;
(y(q)-y(0))
\right)
+\delta (w-1) \exp(-\eps\chi) \ .
\end{equation}
The primitive of this function has a very simple dependence on $y$ and
$\eps$. If we define $Q(w)=\int_w^1 \d w'\; P(w')$ we find:
\begin{equation}
Q(w)=\exp\left(-\eps \int_0^w \d q \; (y(q)-y(0))\right) \label{MAGIC}
\label{Q}
\end{equation}
which is particularly well suited for the extraction of
the function $y(q)$ from numerical simulations. Notice that eq. (\ref{MAGIC}) implies that
$\eps^{-1} \ln(Q(w))$ is $\eps$-independent, which is a non-trivial result.

\section{Examples}

In this section we show how our formula looks like in some cases.
The first example we make is the one of spherical models.
These models are defined by a random Gaussian Hamiltonian
$H({\bf \sigma})$ (${\bf \sigma}=\{\sigma_1,...,\sigma_N\}$),
which have correlation function
\begin{equation}
\overline{ H({\bf \sigma})H({\bf \tau})} = N f(q({\bf \sigma},{\bf
\tau}))
\end{equation}
and the spins are subject to the spherical constraint $\sum_i
\sigma_i^2=N$.
In these models, the function $y(q)$ is temperature independent and
equal to 
\cite{theo}
\begin{equation}
y(q) = 1/2f'''(q)/(f''(q)^{3/2})\ .
\end{equation}

This last equation makes sense only if the
resulting function $y(q)$ is an increasing function of $q$.
This in particular happens for $f(q)=1/2 (q^2+a q^p)$
if $p\geq 4$ and $a$ small enough, where
we find
\begin{equation}
y(q)=\frac{a}{\sqrt{2}}\frac{1}{(p-3)!}\frac{q^{p-3}}
{\left( 2+\frac{a}{(p-2)!}q^{p-2}\right)^{3/2}}\ ,
\end{equation}
while
\begin{equation}
Y(w)\equiv\int_0^w \d q y(q) = 1-\frac{1}{\sqrt{1+\frac{a q^{p-2}}{2 (p-2)!}}}\ .
\end{equation}

\begin{figure}
\epsfxsize=120pt \epsffile{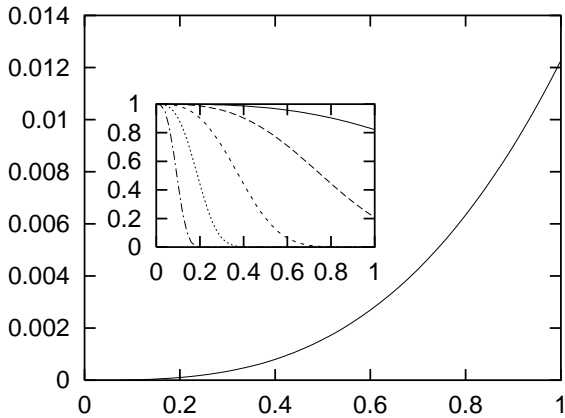}
\caption[0]{\protect\label{due}
The function $Y(w)$ in the spherical model with $p=5$ $a=0.3$. 
In the inset, the function $Q(w)$ for $\eps=2^k$, from top to 
bottom $k=4,7,10,13,16$}
\end{figure}

\begin{figure}
\epsfxsize=200pt \epsffile{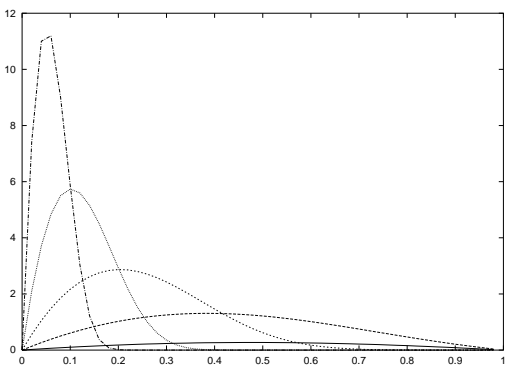}
\caption[0]{\protect\label{duebis}
The function $P(q)$ for the same model and parameters of figure \ref{due}.}
\end{figure}

In figure \ref{due}
we show the 
function $Y(w)$ and the function $Q(w)$ for various values of $\eps$ 
in the case $p=5$, $a=0.3$. In figure \ref{duebis} we show the 
function $P(q)$ for the same values of the parameters. 

In the case of the SK model the function $y(q)$ at low temperature,
has be estimated in \cite{PaT}
using the so-called PaT approximation, and 
displays a square root
divergence at $q=1$, while starting linearly at $q=0$ (a best fit of 
the form $y(q)=\frac{a q+b q^2}{\sqrt{1-q}}$
gives $a=1.309$, $b=-0.695$). 

\begin{figure}
\epsfxsize=200pt \epsffile{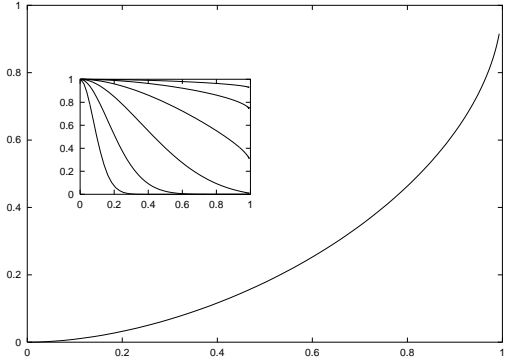}
\caption[0]{\protect\label{sk}
The function $Y(w)$ for the SK model in the PaT approximation.
In the inset, the function $Q(w)$ computed by (\ref{Q}) for
$\eps=0.01 \times 2^k$ with (from top to bottom) $k=3,5,7,9,11,13$.}
\end{figure}

Using that estimate we immediately compute the function $Q(w)$ which
is plotted in figure \ref{sk} for various values of $\eps$.

It is interesting to study the limit $\epsilon \to \infty$
of our formula.
Let us suppose that $y$ behaves as 
$y(q)=a q^\alpha +....$ for $\alpha>1$ for low $q$. In this case $P(q)$ will
be dominates by the behavior of $y$ close to $q=0$. We can introduce
a cut-off $\Lambda$ such that $\Lambda\to 0$ and $\Lambda^{1+\alpha}
\eps \to\infty$.
\begin{equation}
P(q)\simeq \epsilon\alpha q^{\alpha-1} \int_q^\Lambda
\exp\left(-\frac{\epsilon}{\alpha+1}q^{\alpha+1}\right)
\end{equation}
which, rescaling the integration variable and sending the
cut-off to zero becomes:
\begin{equation}
P(q)=\frac{1}{q}
\left( \frac{\epsilon a q^{\alpha+1}}
{1+\alpha}\right)^{\frac{\alpha}{1+\alpha}}
\Gamma (\frac{1}{1+\alpha},\frac{\epsilon a}
{\alpha +1}q^{\alpha+1})
\end{equation}
where the incomplete gamma function is
\begin{equation}
\Gamma(n,x)=\int_x^\infty dy\; y^{n-1}\; e^{-y}.
\end{equation}
This case is relevant in the spherical models where
$y(q)\sim \frac{a}{4 (p-3)!}q^{p-3}$ and for the SK model where 
$y(q)\sim a q$.

\section{Directed polymers in $1+1$ dimension}

The natural play ground of the exposed theory are finite dimensional
spin glasses. Our analysis predicts a very peculiar dependence of the
probability of $q$ and $w$ on $\eps$.  Formulae (\ref{P},\ref{Q}) has
been derived assuming that the low lying states verify
ultrametricity. It is therefore interesting understand what violations
of the scaling forms (\ref{P},\ref{Q}) can be expected when the ground
states structure is nontrivial, but ultrametricity does not hold.

In this paper we study numerically the simple case of directed
polymers in random media in 1+1 dimension. The model we use is defined
on the square lattice, where the polymer can perform a random walk
starting from the origin. On each site of the lattice is defined an
passage energy cost which is a Gaussian variable with unit variance,
and independent of all the other energies.  The properties of this
sort of models have been studied extensively \cite{poli}, and it is
well known that while the low temperature thermodynamics is dominated
by a single ground state, there exist many ``pure states'' (i.e.
metastable states separated by growing barriers) with
typical energy gap with the ground state scaling as $L^{1/3}$ (for a
polymer of total length $L$).  The overlap for two polymers of length
$L$ with a common source in the origin is often defined as the
fraction of the monomers passing in the same sites of the lattice in
the two polymers. Given the previously mentioned scaling of the energy
gap it is natural that the scaling of the coupling in order to have a
non trivial $P(q)$ as defined in the previous section is
\begin{equation}
\eps= \eta L^{1/3}.
\label{scaling}
\end{equation}
If we had to suppose the validity of the formula (\ref{P}) we would
conclude that the function $y(q)$ for samples of length $L$ depends on
$L$ and scales as $L^{-1/3}$.

\begin{figure}
\epsfxsize=200pt \epsffile{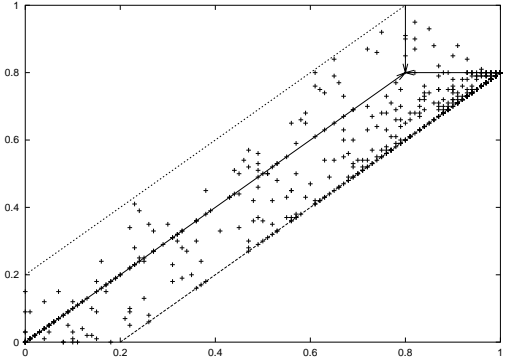}
\caption[0]{
\protect\label{um}
We plot here $q_{23}$ as a function of $q_{13}$ for $q_{12}=0.8$
for 1151 different samples with $L=200$ and
$\eps=0.01$. It is apparent that ultrametricity is violated; if it
was obeyed the points
would stick on the line $q_{23}=q_{13}$ for $q_{13}<0.8$, on the
line $q_{23}=0.8$ for $q_{13}>0.8$, and on the line
$q_{13}=0.8$ for $q_{23}>0.8$. 
The lines $q_{23}=q_{13}\pm 0.2$
represent the bounds imposed by the triangular inequality. Data
obtained for $L=400$ show that there is no trend towards
ultrametricity as $L$ is increased. 
}
\end{figure}

As we stressed, ultrametricity does not hold in this model.
Although we did not make a systematic study, we can easily show
the lack of ultrametricity generalizing the coupling procedure to
a third ``replica'', which has a repulsion both with the unperturbed
ground state, and with the one obtained with the coupling.
For simplicity, we look at the case in which all couplings are equal.
In figure \ref{um}
we show  the overlap between the second and the third replica
as a function of the overlap between the first and the third, fixing
the overlap between the first and the second to 0.8, and we see no
sign of ultrametricity.

\begin{figure}
\label{scaling.eps}
\epsfxsize=200pt \epsffile{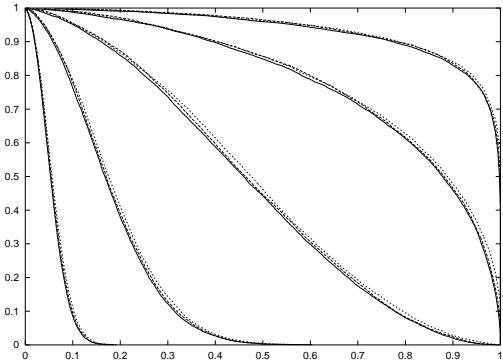}
\caption[0]{\protect
The cumulative probability
$Q(w)$ and 5 different values of $\eta={0.1\times1.4^n
}$, from top to bottom $n=1,5,10,15,20$ and $L=100$ (dotted line) $L=200$
(dashed line) $L=400$ (full line).
The data are obtained on sets of 10000 different samples.
We see that the scaling with L is reasonably obeyed.
}
\end{figure}

In figure \ref{scaling.eps} we show the function
$Q(w)$ for various values of $L$ and $\eps$ of the 
form (\ref{scaling}). 
We see that for values of $L\simeq 200$
the expected independence  of $Q(w)$ of $L$ is reasonably obeyed.
A close inspection to $Q(1)$ however, which represents the probability of
$q=1$ reveals that this quantity behaves as $Q(1)=\exp(-\eta \chi L^{2/3})$,
with $\chi=0.85 \pm 0.02$, and that the scaling is violated
in proximity of $w=1$.

\begin{figure}
\epsfxsize=200pt \epsffile{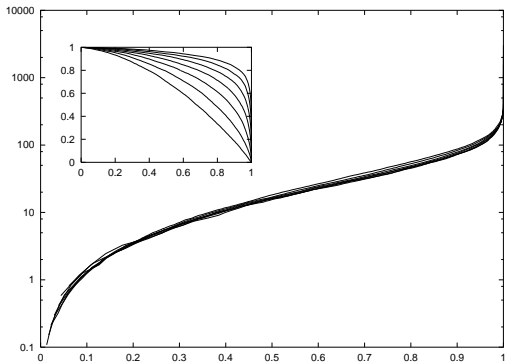}
\caption[0]{\protect\label{piccolo}
The cumulative probability
$-1/\eps Log(Q(w))$ and 7 different values of $\eta={0.1\times1.4^n
}$, $n=1,...,7$ and $L=400$.
The data are obtained on sets of 10000 different samples.
We see that for these low values of $\eps$ the scaling
is reasonably obeyed.
In the inset we plot $Q(w)$ for the same values of the
parameters.}
\end{figure}

\begin{figure}
\epsfxsize=200pt \epsffile{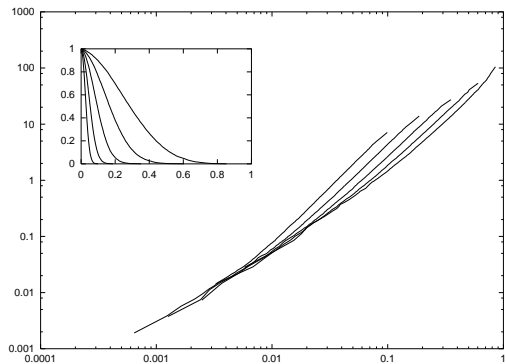}
\caption[0]{\protect\label{grande}
The cumulative probability
$-\eps^{-1} \ln(Q(w))$ and 5 different values of $\eta={0.1\times1.4^n}$,
$n=15,...20$ and $L=400$.
The data are obtained on sets of 10000 different samples.
In the inset the function $Q(w)$.
For these values of $\eps$ the scaling form of the function $Q(w)$ is
violated.
}
\end{figure}

We next try to scale the data with $\eps$ according to the
formula (\ref{Q}). In figure \ref{piccolo} we see that the works
quite well for small values of $\eps$, while we show in figure
\ref{grande} that there are important violations for large values of
$\eps$.

\section{Conclusions}

Monte-Carlo simulations of finite dimensional spin glasses,
show a behavior in agreement with RSB \cite{ybook}. However,
the use of Monte-Carlo has been criticized on the ground that
one can only equilibrate the system too close to the critical point,
where finite size effects are large and could spoil the
conclusions about ergodicity breaking in the thermodynamic limit
 \cite{moore} (see however \cite{NOI,ESSI}).
It is important therefore to find consequences of RSB at zero
temperature.
In this paper we have devised some of them.

We have found that a universal formula holds for the probability of
the overlap among the uncoupled ground state and the coupled one.  The
investigation of the validity of that formula in three dimensional
systems is not beyond reach with the present technology, and will
furnish an important test to understanding the nature of the
spin-glass phase of three dimensional systems.

Two main ingredient will be involved in the calculation: the
exponential distribution of the states \cite{mpv} and their
ultrametric organization \cite{mpstv}. This implies that the tree of
states is described by a single function $y(q)$.  The function $y(q)$
may in principle depend on $N$. Mean field theory predicts that $y(q)$
remains finite in the thermodynamic limit implying that the energy
differences between pure states remain finite in the thermodynamic
limit. However, one could envisage systems where both exponential
distribution and ultrametricity are valid, but the typical energy
differences scale as $L^\theta$. In this case, one still have a
function $y(q)$ which scales as $L^{-\theta}$ and in order to measure
a nontrivial overlap distribution one needs a coupling of order
$L^\theta$. The numerical study of the overlap and gap distribution, and the
comparison with the formulae found in this paper will give important
information about the organization of the states in short range
models.

\section*{Acknowledgments}

We thank Enzo Marinari, Matteo Palassini and Peter Young
for important discussions.


\begin{thebibliography}{99}

\bibitem{MMPRRZ} E. Marinari, G. Parisi, F. Ricci-Tersenghi, J.J. Ruiz-Lorenzo and F.
Zuliani, J. Stat. Phys. (2000)

\bibitem{PY} M. Palassini and P. Young Phys.Rev.Lett. 83 (1999)
5126-5129, Phys.Rev. B60 (1999) R9919

\bibitem{PY2}  M. Palassini and P. Young Preprint cond-mat/0002134 

\bibitem{HM}  Houdayer and O. Martin
 Phys. Rev. Lett. 82 (1999) 4934-4937

\bibitem{KM} F. Krzakala, O. C. Martin, cond-mat/0002055 {\it Trivial link but non-trivial spin 
overlaps in 3-dimensional spin glasses} 


\bibitem{PAL} M. Palassini, {\sl Tesi di Perfezionamento, Scuola
Normale Superiore}, Pisa (2000).

\bibitem{MP} E. Marinari and G. Parisi, cond-mat/0002457 
{\sl Comment on ``Triviality of the Ground State Structure in Ising
Spin 
Glasses''}, 
{\sl On the Effects of Changing the Boundary Conditions on the 
Ground State of Ising Spin Glasses} 
cond-mat/0005047 and work in 
progress.

\bibitem{repequiv} G. Parisi, preprint cond-mat/9801081
 S. Franz, M. Mezard, G. Parisi, L. Peliti 
J. Stat. Phys. {\bf 97} (1999) 459, G. Parisi and  F. Ricci-Tersenghi
J. Phys. A {\bf 33}, 113 (2000).

\bibitem{MPV}
M.~M{\'e}zard, G.~Parisi and
M.~A.~Virasoro, \textit{Spin Glass Theory  and Beyond} (World
Scientific, Singapore, 1987)

\bibitem{PaT}
G. Parisi and  G. Toulouse
J. Physique (Paris), Lettres, {\bf 41}, (1980) L361,
J. Vannimenus, G. Toulouse, G. Parisi,
J. de Physique I, {\bf 42}, (1981) 565

\bibitem{theo} Th. M. Nieuwenhuizen
Phys. Rev. Lett. {\bf 74}, 4293 (1995)

\bibitem{poli} M. Mezard, J. Physique 51 (1990) 1831, M. Mezard and
G. Parisi, J.Phys. A25 (1992) 4521.

\bibitem{ybook}
E. Marinari, G. Parisi and
J.~J.~Ruiz-Lorenzo in
A. P. Young (ed.), \textit{Spin Glasses and Random Fields}
(World Scientific, Singapore, 1997).

\bibitem{moore} M. A. Moore, H. Bokil and B.Drossel, Phys. Rev. 
Lett. {\bf 81},
4252 (1998).
B. Drossel, H. Bokil, M. A. Moore, and 
A. J. Bray,
European Physical Journal B {\bf 13}, 369-375 (2000).

\bibitem{NOI} E. Marinari, G. Parisi, J.J. Ruiz-Lorenzo and F. Zuliani, Phys. Rev. Lett.
{\bf 82}, 5176 (1999); M. A. Moore, H. Bokil and B.Drossel, Phys. Rev. Lett. {\bf 
82}, 5177 (1999).

\bibitem{ESSI} F. Ricci-Tersenghi and F. Ritort, J. Phys. A {\bf 33}, 3727 (2000).
\bibitem{mpv}
M.~M{\'e}zard, G.~Parisi and
M.~A.~Virasoro, J. Phys. Lett. {\bf 46} (1985) L217;
Europhys. Lett. {\bf 1} (1986) 77.

\bibitem{mpstv} M. M\'ezard, G. Parisi, N. Sourlas, G. Toulouse
and M.A. Virasoro, J. Physique {\bf 45} 843 (1985).


\end{thebibliography}
\end{document}